\font\tx=cmr10 at 11pt     \font\it=cmti10 at 11pt
\font\ma=cmmi10 at 11pt    \font\sy=cmsy10 at 11pt
\textfont0=\tx        \textfont1=\ma         \textfont2=\sy
\font\sub=cmr8        \font\masub=cmmi8      \font\sysub=cmsy8
\scriptfont0=\sub     \scriptfont1=\masub    \scriptfont2=\sysub
\font\title=cmbx12    \font\author=cmr12     \font\bx=cmbx10 at 11pt
\baselineskip=13pt    \parindent=18pt        \raggedbottom
\lefthyphenmin=3      \righthyphenmin=4      \hyphenpenalty=200
\def\space{\vskip 13pt}
\def\section#1{\space\space\goodbreak\leftline{\bx#1}\nobreak\space}
\vglue 5mm
\tx
\centerline{\title  IMPLICATIONS OF BINARY PROPERTIES FOR}
\smallskip
\centerline{\title        THEORIES OF STAR FORMATION
          \footnote{$^1$}{\tx Presented at IAU Symposium 200, ``The
         Formation of Binary Stars'', held in Potsdam, Germany, April
          10--15, 2000; to be published by the Astronomical Society
          of the Pacific, edited by R. D. Mathieu and H. Zinnecker}}
\space\space

\centerline{\author         Richard B. Larson}
\space
\centerline{            Yale Astronomy Department}
\centerline{          New Haven, CT 06520-8101, USA}
\centerline{               larson@astro.yale.edu}
\space\space\space

\centerline{                    ABSTRACT}
\space
{\narrower
   The overall frequency and other statistical properties of binary
systems suggest that star formation is intrinsically a complex and
chaotic process, and that most binaries and single stars actually
originate from the decay of multiple systems.  Interactions between
stars forming in close proximity to each other may play an important
role in the star formation process itself, for example via tidally
induced accretion from disks.  Some of the energetic activity of newly
formed stars could be due to bursts of rapid accretion triggered by
interactions with close companions.
\space}

\section{1.~~Introduction}

   What can the study of binary stars tell us about how stars form?
Stars and stellar systems can be viewed as fossils that preserve some
record of the star formation process, and we wish to understand what
they can tell us about this process.  Single stars have only one
parameter that is conserved from their time of formation, namely their
mass, so the main thing we can hope to learn about star formation from
single stars is the stellar initial mass function, which may constrain
the origin of stellar masses.  Binary systems have three additional
conserved parameters, namely their angular momentum, eccentricity,
and mass ratio, and therefore the statistical  properties of binaries
can in principle place stronger constraints on the nature of the star
formation process.  This is especially true concerning the small-scale
dynamics of star formation and the mechanisms by which matter actually
becomes incorporated into stars.

   Most fundamentally, the frequency of binary and multiple systems
tells us that most if not all stars are formed in such systems and not
in isolation.  The `standard model' that has been developed to describe
the formation in isolation of single stars like the Sun therefore cannot
apply to most stars.  Clearly the Sun is not typical in being a single
star, but even the Sun could have formed in a multiple system because
the plane of its planetary system is tilted with respect to the solar
equatorial plane, plausibly because of a close encounter with another
star soon after the Sun was formed (Herbig \& Terndrup 1986; Heller
1993).  Therefore, not only is the formation of binary and multiple
systems clearly nature's preferred way of making stars, it might even
be nature's {\it only} way of making stars.  This possibility would be
appealing theoretically because if much of the angular momentum of a
collapsing cloud goes into  the orbital motion of a binary or multiple
system, this would go a long way toward solving the classical `angular
momentum problem' of star formation.

   A further basic fact about binary stars is that their orbital
parameters vary over enormous ranges and show no clearly preferred
values.  This means that no `standard model' of binary formation
with any typical set of parameter values can adequately describe the
formation of most binaries, or of most stars.  A major conclusion to
be elaborated below is that star formation must be, at least to some
extent, intrinsically a chaotic process, involving complex dynamics
and interactions in systems of forming stars, and that statistical
approaches are needed to deal with the broad distribution of outcomes
arising from such processes.  Chaotic dynamics and protostellar
interactions in forming binary and multiple systems may also have
important implications for the nature of the accretion processes by
which stars acquire most of their mass, and may account for some of
the highly variable energetic activity observed in newly formed stars.

\section{2.~~Basic Statistical Properties of Binaries}

   Although a clear consensus has not yet been reached on all of the
statistical properties of binaries, the following basic properties seem
reasonably clear:

\medskip\noindent
(1) {\it Frequency:}  The overall frequency of binaries, defined as the
fraction of primaries that have at least one companion, is at least 50
percent (Heintz 1978; Abt 1983; Duquennoy \& Mayor 1991; Mayor et al.\
1992).  The binary fraction appears to increase with increasing primary
mass, at least among the more massive stars: the O~and~B stars have a
companion frequency of at least 70~percent (Abt, Gomez, \& Levy 1990;
Mason et al.\ 1998; Preibisch et al.\ 1999; Preibisch 2001), while for
G~stars the binary frequency is around 50~percent (Duquennoy \& Mayor
1991; Fischer \& Marcy 1992) and the M~stars may have an even lower
binary frequency of around 30--40 percent (Fischer \& Marcy 1992;
Tokovinin 1992; Mayor et al.\ 1992).  Brown dwarfs are rare as
companions to lower-main-sequence stars, although brown-dwarf binaries
appear not to be rare (Basri 2001).  An increase in binary frequency
with mass would be expected if most stars form in multiple systems that
disintegrate, since the more massive stars would then preferentially
remain in binaries while the less massive ones would preferentially be
ejected as single stars.  The binary frequencies summarized above and
their dependence on mass are in fact consistent with the results of
simulations of the decay of small multiple systems (Sterzik \& Durisen
1998, 1999), and therefore they are consistent with the possibility
that all stars are formed in such systems.

\medskip\noindent
(2) {\it Period Distribution:}  The periods of binaries are distributed
continuously over an extremely large range (Heintz 1978; Abt 1983;
Griffin 1992), and in a frequently quoted study Duquennoy \& Mayor
(1991) found a broad and nearly flat distribution in the logarithm of
the period which they fitted with a gaussian function centered on a
median period of 180 years, corresponding to a median semi-major axis of
about 35~AU.  Pre-main-sequence stars show a very similar distribution
of periods and separations (Mathieu 1994; Simon et al.\ 1995).  The most
remarkable feature of this distribution is its flatness, i.e.\ the fact
that the number of systems per unit logarithmic interval is almost
constant over many orders of magnitude in period or separation.  Other
authors have made the same point by noting that the distribution of
semi-major axes $a$ follows the power-law form $f(a)\propto a^{-1}$,
equivalent to a flat distribution in $\log a$, over many orders of
magnitude in $a$ (Heacox 1998, 2000; Stepinski \& Black 2000a,b).  This
distribution is nearly scale-free and implies that there is no strongly
preferred scale for the formation of binary systems.

\medskip\noindent
(3) {\it Eccentricities:}  Binaries with periods longer than a year,
which are not significantly affected by tidal circularization, have a
broad distribution of orbital eccentricities $e$ that is nearly flat for
$0 < e < 1$ (Aitken 1935; Duqennoy \& Mayor 1991; Mayor et al.\ 1992),
with a median value of around 0.55.  Clearly there is no tendency for
binaries to form with nearly circular orbits, and high eccentricities
are common; thus models of binary formation that postulate nearly
circular orbits cannot adequately describe the formation of most
systems.

\medskip\noindent
(4) {\it Mass Ratios:}  The distribution of mass ratios $q = M_2/M_1$
has been the most difficult function to pin down because it is subject
to many biases and selection effects, and because it depends on period
and probably also on primary mass.  The strongest conclusion seems to
be that the distribution of $q$ values is different for short-period
and long-period binaries: according to Abt \& Levy (1978) (see also
Abt 1983 and Abt \& Willmarth 1992), for systems with periods less
than about 100~years (i.e., semi-major axes less than about 25~AU),
the distribution of $q$ values is much flatter than would be predicted
if the stars had been randomly selected from a standard IMF, while for
longer-period systems the distribution of $q$ is more consistent with
what would be predicted from the IMF.  Mayor et al.\ (1992) and Mayor
(2001) confirm that spectroscopic binaries have a distribution of $q$
values that is nearly flat in the range $0.2<q<1$, implying that the
masses of the stars in these systems tend to be more nearly equal than
would be predicted by random selection from the IMF.

\medskip
   As we have heard from Latham (2001), on the basis of impressive
statistics, the statistical properties of binaries in the Galactic halo
are in all respects indistinguishable from those of the binaries in the
Galactic disk.  Thus these properties are of great generality, and are
not restricted to any particular place or time of formation.

\section{3.~~Implications for Theories of Star Formation}

   From the above brief summary of the basic statistical properties of
binaries, we can draw the following inferences for theories of star
formation:

\medskip\noindent
(1) At least two-thirds of all stars are in binary or multiple systems,
and this can only be a lower limit to the fraction of stars formed in
such systems.  The statistical evidence summarized above is consistent
with the possibility that {\it all stars are formed in binary or
multiple systems, and the minority of single stars result from the decay
of multiple systems,} as suggested by Heintz (1969) and Larson (1972).
As an example, if stars typically form in triple systems that decay into
a binary and a single star, this would yield similar numbers of binaries
and single stars, as observed.  Numerous simulations of the collapse and
fragmentation of dense cloud cores, including many presented at this
meeting, suggest that such multiple fragmentation processes are a very
general result (Bodenheimer et al.\ 2000; Bodenheimer 2001; Bonnell
2001; Boss 2001; Klein 2001; Whitworth 2001).  If so, {\it a separate
mechanism for forming single stars is not required.}

\medskip\noindent
(2) Given the above median orbital parameters, a typical star forms with
a companion in an orbit having a period of about 180 years, a semi-major
axis of about 35~AU, and an eccentricity of about~0.55.  The period and
size of this `median orbit' are similar to those of the planet Neptune,
but this orbit is quite eccentric, unlike that of Neptune, and the
separation of the two stars varies from about 16 to 54~AU.  The
formation of a planetary system similar to our own is clearly not
possible such a situation, and {\it any remaining circumstellar disk
will be strongly disturbed by the tidal effect of the companion} at
every periastron passage.  Circumstellar disks may be even more
strongly perturbed for stars that form in multiple systems.

\medskip\noindent
(3) Since the orbital parameters of binaries vary widely, the detailed
circumstances of star formation will also vary greatly from case to
case, and most stars will not form in circumstances very similar to the
above `typical' case.  Thus, {\it there can be no `standard model' for
binary formation,} or for the formation of stars generally, and a more
statistical approach to the problem is needed.  For example, instead of
continually refining the accuracy of simulations of one or a few special
cases, it may be more useful for future theoretical work to explore with
less precision a larger parameter space and to try to predict the
statistical distribution of outcomes.

\medskip\noindent
(4) Since the stars in systems with separations smaller than a few tens
of AU tend to have masses that are more nearly equal than would be
predicted if they had been randomly selected from the IMF, {\it the
masses of stars that form within a few tens of AU of each other are
correlated.}  This means that the mechanisms that determine stellar
masses cannot be purely local to the star, and that effects acting on
scales at least as large as a few tens of AU must play a role.  The
observed correlation could not be accounted for if, for example, stars
``determine their own masses'' by feedback effects that are purely local
to the star and that act independently of the larger-scale environment.
Stars must know something about the environment in which they form.

\medskip\noindent
(5) The fact that the more massive stars tend to have more numerous
and more massive companions suggests that {\it interactions with
companions play an increasingly important role in the formation of the
more massive stars.}  It has been suggested that accretional processes
associated with interactions, perhaps even including direct stellar
collisions and coalescence, may play important roles in the formation
of the most massive stars (Bonnell, Bate, \& Zinnecker 1998; Stahler,
Palla, \& Ho 2000) and in the origin of the upper IMF (Larson 1999;
Bonnell 2000), and these suggestions receive support from the observed
higher frequency of companions among massive stars.

\medskip
   The following sections will consider further some of the above
implications of the statistical properties of binaries for star
formation, especially the effects of tidal interactions on disks in
binary systems and the role of the chaotic dynamics of systems of
forming stars in explaining the broad distribution of binary orbital
parameters.

\section{4.~~Effect of Companions on Disk Evolution}

   Numerical simulations of star formation often produce circumstellar
disks, and remnant disks are also observed to be common around newly
formed stars, even in binary systems (Mathieu 1994; Mathieu et al.\
2000).  A circumstellar disk in a binary system whose separation is not
much larger than the size of the disk will be strongly tidally perturbed
by the companion every time it passes periastron, and simulations show
that these perturbations generate strong two-armed trailing spiral
structure in the disk (Bate 2000, 2001; Nelson 2000; Nelson, Benz,
\& Ruzmaikina 2000).  If a tidally perturbed disk is continually
replenished with new material, for example from an infalling envelope,
the size and mass of the disk remain roughly constant and the amplitude
of the tidally generated spiral pattern therefore also remains roughly
constant in time, although the form of the pattern continually
fluctuates.  Such tidally produced spiral patterns are at least partly
wave-like in nature, and they tend to propagate inward and dissipate
in the inner part of the disk.

   Tidally generated spiral waves may play an important role in driving
accretion flows in disks (Spruit et al.\ 1987; Larson 1989; Savonije,
Papaloizou, \& Lin 1994).  A trailing spiral wave propagating into a
disk has negative angular momentum, and thus it temporarily reduces the
angular momentum of the disk; if the wave is somehow dissipated, the
angular momentum of the disk is permanently reduced, and this can
drive an inflow.  For strong waves, a likely dissipation mechanism
is the formation of shocks, and it was first suggested by Shu (1976)
that spiral shocks might drive accretion flows in disks.  Numerical
simulations showing that tidally generated spiral shocks can indeed
drive strong inflows in disks were presented by Sawada, Matsuda, \&
Hachisu (1986) and Sawada et al.\ (1987), and self-similar solutions
for shock-driven accretion were obtained by Spruit (1987).  The disks
studied by Sawada et al.\ (1986, 1987) were two-dimensional and had
unrealistically high temperatures, but spiral shocks are also found
in recent three-dimensional simulations, where these shocks are more
tightly wound and resemble those inferred to exist in some cataclysmic
variable systems (Haraguchi, Boffin, \& Matsuda 1999; Makita, Miyawaki,
\& Matsuda 2000; Matsuda et al.\ 2000).

   Accretion flows driven by tightly wound waves are a relatively weak
effect, and therefore they are difficult to study numerically.  Wave
theory may then prove useful, and wave profiles for non-linear acoustic
waves in disks have been calculated by Larson (1990a) in order to
determine the associated accretion rates.  As with water waves, the wave
profile is sinusoidal for small amplitudes but becomes increasingly
sharply peaked at its crest as the amplitude increases, eventually
`breaking' to form a discontinuity or shock when the amplitude exceeds
a critical value.  Density profiles of the predicted form with
inward-propagating shocks are seen in the high-resolution simulations
of R\'o\.zyczka \& Spruit (1993), which also show that the resulting
wave pattern is often complex and time-dependent.  The value of the
Shakura-Sunyaev alpha parameter for a steady spiral wave pattern
containing shocks can be estimated from the results of Larson (1990a)
to be of the order of $(3$--$10) \times 10^{-4}$ for disks like
those discussed here, implying an inflow timescale of the order of
$(2$--$6) \times 10^5$ years at a radius of $1\,$AU\null.  The inflow
rate found numerically by R\'o\.zyzcka \& Spruit (1993) corresponds
to an average alpha of about $10^{-3}$, which is sufficient to drive
significant inflows in circumstellar disks.

   A full understanding of transport processes in disks does not yet
exist, and several mechanisms could be involved.  Wave effects will be
important at least in the outer tidally disturbed parts of disks like
those discussed here, where strong spiral waves are present.  A trailing
spiral wave pattern creates a gravitational torque that also acts to
transport angular momentum outward (Lynden-Bell \& Kalnajs 1972; Larson
1984), but if the disk is of low mass and gravitationally stable, as
most observed disks appear to be, the wave transport effect is more
important than the gravitational torque for the same spiral pattern
(Larson 1989).  The ultimate fate of tidal waves propagating into a
circumstellar disk is not yet clear, and it depends on how and where
they are dissipated.  If the waves are strongly damped at some radius,
inflowing matter may tend to pile up there unless other transport
mechanisms take over.  If no other effect becomes more important, the
continuing accumulation of matter at any radius will eventually cause
gravitational instability to occur and gravitational torques to become
important (Larson 1984; Stahler 2000), so inflow of matter seems likely
to continue through one mechanism or another even if the tidal waves do
not propagate all the way to the center.

   Tidal effects can be particularly strong in very eccentric binaries,
where they may cause episodes of rapid accretion at each periastron
passage; this is suggested by the simulations of Bonnell \& Bastien
(1992) in which violent spiral disturbances are generated at each
periastron passage and lead to bursts of rapid accretion.  Several
transport mechanisms may play a role in these accretion events, since
in addition to generating the wave effects discussed above, strong tidal
disturbances may also result in enhanced gravitational and turbulent
transport effects (Nelson et al.\ 2000).  Even more violent disturbances
that disrupt disks and/or trigger bursts of rapid accretion may be
produced by close encounters in multiple systems, as is again
illustrated by simulations of such encounters (Heller 1995; Boffin et
al.\ 1998; Pfalzner, Henning, \& Kley 2000).  If interactions with close
companions can indeed trigger episodes of rapid accretion onto forming
stars, this could help to explain the intriguing associations that have
been reported at this meeting between the presence of close companions
and the occurrence of protostellar jets (Reipurth 2001) and extreme
T~Tauri activity (Mathieu 2001).  The FU~Orionis phenomenon may also
be explainable in a similar way (Bonnell \& Bastien 1992).  A possible
analogy may be noted between these phenomena of early stellar evolution
and nuclear activity in galaxies, since the most extreme forms of
activity in galaxies are also caused by violent tidal interactions
and mergers that drive strong inflows in disks (Larson 1994).

\section{5.~~Understanding the Broad Period Distribution}

   The fact that binaries are distributed in a logarithmically nearly
uniform way over many orders of magnitude in period and separation is
a challenge to theories of star formation, since standard models would
tend to predict characteristic values of these quantities, while the
observations show no evidence for any preferred scales and suggest
instead a nearly scale-free formation process.  The scale-free nature
of the distribution of separations is also evident from the form of
the distribution of semi-major axes, which in linear units is $f(a)
\propto a^{-1}$ (Heacox 1998, 2000; Stepinski \& Black 2000a,b),
and from the dependence on angular separation of the average surface
density of companions on the sky, which follows $\Sigma(\theta) \propto
\theta^{-2}$ (Larson 1995; Simon 1997).  In the latter representation
of the data, it is striking that the companion surface density falls
off much more steeply with separation for binaries than it does for
the larger-scale clustering of young stars in regions of star formation
(Larson 1995; Bate, Clarke, \& McCaughrean 1998).  This suggests that
different processes act on different scales to determine the spatial
distribution of young stars, and that processes acting on the scale of
binaries produce a large excess of close pairs compared with what would
be predicted from an extrapolation of the larger-scale clustering of
young stars.

   This distinction between binary systems and the larger-scale
clustering of young stars is reminiscent of the situation for galaxies,
which are also much more compact in structure than would be predicted
from their clustering properties.  There is even a quantitative
similarity between the distribution of separations of binaries and the
spatial distribution of stars in elliptical galaxies: as was first found
by Hubble (1930; see also Holmberg 1975), the surface brightnesses of
elliptical galaxies fall off approximately as the inverse square of
distance from the center, and this resembles the inverse-square
dependence of the surface density of binary companions on separation.
Although the light profiles of elliptical galaxies tend to become
shallower with increasing luminosity (Schombert 1987), a power law of
slope $-\,2$ remains a representative approximation for galaxies of
intermediate luminosity.  The stars in binary systems and the stars
in elliptical galaxies are thus both distributed roughly uniformly with
respect to the logarithm of separation or distance from the center of
mass.  Similar surface density profiles have been found for some young
star clusters (Moffat, Drissen, \& Shara 1994), so it may be a general
phenomenon that systems of stars tend to be formed with their stars
distributed roughly uniformly in the logarithm of separation.

   The centrally condensed structures of elliptical galaxies are
believed to result from dissipative formation processes, which could
involve either gaseous dissipation or dynamical friction effects or
both (Kormendy 1990; Larson 1990c).  The fact that binary systems are
relatively tightly bound compared to larger groupings of young stars
suggests that dissipative effects have played a role in their formation
too (Larson 1997; Heacox 1998, 2000).  Such effects generally involve
the loss of both energy and angular momentum, but since the energy of
a collapsing cloud is easily radiated away while its angular momentum
is less easily disposed of, the angular momentum is probably the most
important dynamical parameter determining the outcome of the collapse
and the separation of any resulting binary.  The median specific angular
momentum of binary systems is about an order of magnitude smaller than
that of dense cloud cores, so the formation of a typical binary from
a typical cloud core must involve a reduction in specific angular
momentum by about an order of magnitude (Bodenheimer 1995).  In
addition, the amount of angular momentum lost or redistributed during
the formation process must also vary widely from case to case to
account for the observed broad distribution of binary separations.

   If the period distribution found by Duquennoy \& Mayor (1991) is
approximated by the gaussian function of the logarithm of the period
suggested by these authors, the corresponding distribution of the
specific angular momentum $j$ of binaries is a gaussian function of
log$\,j$ with a mean of about $-\,3.6$ and a width at half maximum of
about 1.8 if $j$ is measured in km$\,$s$^{-1}\,$pc and logarithms to
the base 10 are used.  This is a much broader distribution than would
be produced by any simple gaussian random process, since such a process
would tend to generate a distribution that is gaussian in each of the
three components of $j$ rather than one that is gaussian in log$\,j$.
For example, if each of the three components of $j$ has a gaussian
distribution with a mean of zero and a standard deviation $\sigma$, the
distribution of log$\,j$ is proportional to $j^3\,$exp($-j^2/2\sigma^2$)
and has a width at half maximum of only 0.43.  The median initial value
of log$\,j$ suggested by the velocity gradients measured for cloud cores
by Goodman et al.\ (1993) is about $-\,$2.7, so if cores like these are
to form binary systems with the observed distribution of properties, the
median log$\,j$ must be reduced from $-\,$2.7 to~$-\,$3.6.  At the same
time, any plausible initial distribution of log$\,j$ must be broadened
considerably; in the above example, it must be broadened by more than
a factor of~4 from a width of 0.43 to a width of 1.8 at half maximum.

   Since the mechanisms that form binaries must accordingly reduce
log$\,j$ by an amount that varies widely from case to case, we can
regard the amount by which log$\,j$ is reduced as a random variable
which has a large dispersion.  For example, if we denote by $X$ the
amount by which the natural logarithm ln$\,j$ is reduced from its
initial value ln$\,j_0$, we can write $j = j_0 e^{-X}$ where $X$ is a
random variable whose mean and standard deviation must be about 2.1 and
1.7 respectively to account for the observations.  Such an exponential
dependence of a physical quantity on a negative exponent is suggestive
of a damping or decay process; a familiar astronomical example is
interstellar extinction, whereby a moderate dispersion in optical depth
$\tau$ produces a large dispersion in the apparent brightnesses of
stars, which vary as $e^{-\tau}$.  To illustrate how such an effect
might operate on a dynamical variable like $j$, suppose that the angular
momentum of a forming binary system is reduced by a decelerating torque
or drag effect, which might be of gravitational, magnetic, or viscous
origin, and suppose that its angular momentum has an associated decay
rate $A$ such that $dj/dt = -Aj$; then after any time $t$ we have
$j = j_0 e^{-At}$, so that if either the magnitude $A$ of the drag
effect or the time $t$ over which it operates varies significantly
from case to case, a large logarithmic dispersion in specific angular
momentum $j$ can be produced (Larson 1997; Heacox 1998).

   Angular momentum is conserved only in an axisymmetric system, and
gravitational torques will redistribute angular momentum whenever there
are departures from axial symmetry; for example, the gravitational drag
that acts on orbiting clumps in a fragmenting cloud tends to reduce
their orbital angular momentum and cause them to form more tightly bound
systems (Larson 1978, 1984).  Boss (1984) showed that this effect can
in some cases reduce the angular momentum of a forming binary system by
a large factor within an orbital period (see also Boss 1988, 1993 and
Bodenheimer 1995).  This gravitational drag effect is closely analogous
to the `dynamical friction' of stellar dynamics (Binney \& Tremaine
1987), and it may play an important role not only in the formation of
binary systems but also in the formation of clusters like the Trapezium
cluster, which has at its center the compact and massive Trapezium
multiple system (Larson 1990b; Zinnecker, McCaughrean, \& Wilking 1993).

   Gravitational forces also vary with space and time in a system with
a clumpy mass distribution, and a random element is introduced if the
dynamics of the system becomes chaotic.  Chaotic dynamics is indeed
expected if stars typically form in multiple systems, as is suggested
by many numerical simulations of collapsing and fragmenting clouds,
including those of Burkert, Bate, \& Bodenheimer (1997; see also Bodenheimer et al.\ 2000) and those presented at this meeting by
Bodenheimer, Bonnell, Boss, Klein, and Whitworth.  In the limit where
most of the mass has condensed into stars, close encounters between the
stars in a forming multiple system will produce large perturbations in
their orbital motions, and these perturbations will on the average
tend to make the closer subsystems more tightly bound, as happens with
binaries in star clusters (Heggie 1975).  A young binary system may in
this situation lose most of its angular momentum through a few close
encounters with other stars.  The overall loss in angular momentum
implied by the median initial and final values of~$j$ given above is
a factor of~8, and this could be achieved through a few encounters if,
for example, each encounter reduces the angular momentum of the system
by a factor of~2 and the average number of such encounters is~3.
Random fluctuations in this small number of events could then account
for much of the dispersion in the final orbital properties of binaries.
The standard deviation in $X$ arising just from the square root of the
number of events is~1.2, but if different encounters vary in their
effects by a similar amount, the total standard deviation in $X$
becomes about 1.7, as required.

   To test whether such effects can account for the observed
distributions of properties of binaries, numerical simulations of
the formation of small multiple systems are needed that predict the
properties of the resulting binaries, but most calculations have not
been carried to the point where most of the mass is in stars, and
therefore they do not yet predict the properties of the resulting
systems.  Among the few simulations that have been carried this far are
the early crude ones of Larson (1978); although their accuracy is low,
these simulations do include the gravitational effects discussed above,
and these effects play an important role in producing the many binary
and multiple systems that are illustrated for example in Figures 6(a)
and 6(b) of Larson (1978).  The clustering of these objects can be
examined by plotting the average surface density of companions as a
function of separation, as was done for T~Tauri stars by Larson (1995),
and the results are very similar: there is a distinct binary regime
on small scales where the companion surface density falls off much
more steeply with separation than on larger scales, again varying
approximately as the $-\,2$ power of separation (Larson 1997).  Although
the range of separations covered by these simulations is small compared
to that represented by the observations, the fact that these results
resemble the observations so closely suggests that the period
distribution of binaries results from basic and universal features of
gravitational dynamics that are present even in these crude simulations.

   The formation of elliptical galaxies has been modeled in much more
detail than the formation of binary systems, and the results of this
work may also be relevant here if general gravitational mechanisms
are involved (Larson 1997).  Elliptical galaxies are believed to be
formed by mergers of smaller systems, or by the collapse of clumpy
protogalaxies containing substructures that are eventually erased.
In either case, the main effect responsible for producing the final
centrally condensed structure of the system is probably dynamical
friction acting on the densest subunits and causing them to sink toward
the center.  The merger simulations of White (1978), Villumsen (1982),
and Barnes (1992) and the simulations of clumpy collapse by van Albada
(1982) and Katz (1991) all yield similar results that reproduce
approximately the observed structures of elliptical galaxies.  In these
simulations, different mass elements experience widely differing amounts
of dynamical friction and energy loss, and this results in a mass
distribution in which the mass is spread out roughly uniformly in
logarithmic intervals of radius, as observed.

   Similar effects may occur statistically in the much smaller systems
considered here, leading again to a distribution of separations that is
roughly uniform in the logarithm of separation.  In the limit where all
of the mass has condensed into stars and the system has become a small
n-body system, simulations of the decay of small multiple systems such
as those made by Sterzik \& Durisen (1998, 1999) become relevant.  The
results of many such simulations show that n-body dynamics can indeed
create a wide spread in the separations of the resulting binaries,
and that this can account for a good part of the observed spread of
separations.  However, the closest binaries are still not reproduced,
and this suggests that strongly dissipative gas dynamical effects are
needed to form these systems.  Thus a combination of gas dynamical and
stellar dynamical effects is probably required to account for the full
range of binary properties: gas dynamics provides the strong dissipation
via shock formation that is needed to make close binary systems and
individual stars, while the stellar dynamical effects that dominate
when the system becomes very clumpy introduce a chaotic element that
leads to a large dispersion in the final results.

\section{6.~~Summary}

   The statistical properties of binary systems all point to a more
complex and dynamic picture of star formation than that provided by
the standard models for isolated star formation that have dominated
theoretical work so far.  From the frequency of binary and multiple
systems it is clear that stars seldom if ever form in isolation, and
the wide dispersion in binary properties suggests that even binaries
typically do not form in isolation but as parts of larger systems whose
dynamics is complex.  The dependence of binary frequency on mass is
consistent with the possibility that all binaries and single stars
originate from the decay of multiple systems; part of the wide range
in binary properties may then result from the chaotic dynamics of such
systems.  Many simulations of the collapse and fragmentation of dense
cloud cores suggest that the typical outcome is indeed the formation
of multiple systems with chaotic dynamics.  This occurs partly because
realistic initial conditions always have some degree of irregularity
that tends to be amplified during the collapse, and partly because
gravitational and hydrodynamic instabilities generate additional
structure during the collapse which ultimately results in chaotic
behavior.  Future theoretical work on star formation will therefore
have to deal with chaotic systems and be able to predict a statistical
distribution of outcomes.

   This complexity of the dynamics also has implications for the
mechanisms by which gas becomes incorporated into forming stars.  In
standard models, most of the matter that goes into a star is assumed to
be acquired by accretion from a disk.  Disks may also be involved in
the formation of stars in binary and multiple systems, but in this case
tidal effects may play an important role in the accretion process.  In
eccentric binaries and in multiple systems, close encounters may produce
particularly violent tidal effects that lead to bursts of enhanced
accretion, and the stars in such systems might acquire much of their
mass as a result of such discrete accretion events.  Some young stars
show variable activity that might reflect episodes of enhanced accretion
triggered by interactions with companions; the FU~Orionis phenomenon and
protostellar jets might have such origins, and there is evidence that
both protostellar jets and extreme T~Tauri activity are associated with
the presence of close companions.  It will be of great interest to try
to establish by further research whether some of the more dramatic forms
of activity in young stars actually reflect an intrinsically violent
and chaotic star formation process.

\section{References}

{\leftskip=5mm \parindent=-5mm

Abt, H. A. 1983, ARA\&A, 21, 343

Abt, H. A., \& Levy, S. G. 1978, ApJS, 36, 241

Abt, H. A., Gomez, A. E., \& Levy, S. G. 1990, ApJS, 74, 551

Abt, H. A.,, \& Willmarth, D. W. 1992, in Complementary Approaches to
   Double and Multiple Star Research, IAU Colloq.\ 135, eds.\ H.~A.
   McAlister \& W.~I. Hartkopf, ASP Conference Series Vol.\ 32, p.~82

Aitken, R. G. 1935, The Binary Stars (McGraw-Hill, New York; reprinted
   by Dover, 1964)

Barnes, J. E. 1992, ApJ, 393, 484

Basri, G. 2001, in The Formation of Binary Stars, IAU Symp.\ 200, eds.\
   R.~D. Mathieu \& H. Zinnecker (ASP, San Francisco), in press (this
   volume)

Bate, M. R. 2000, MNRAS, 314, 33

Bate, M. R. 2001, in The Formation of Binary Stars, IAU Symp.\ 200,
   eds.\ R.~D. Mathieu \& H. Zinnecker (ASP, San Francisco), in press
   (this volume)

Bate, M. R., Clarke, C. J., \& McCaughrean, M. J. 1998, MNRAS, 297, 1163

Binney, J., \& Tremaine, S. 1987, Galactic Dynamics (Princeton
   University Press, Princeton)

Bodenheimer, P. 1995, ARA\&A, 33, 199

Bodenheimer, P. 2001, in The Formation of Binary Stars, IAU Symp.\ 200,
   eds.\ R.~D. Mathieu \& H. Zinnecker (ASP, San Francisco), in press
  (this volume)

Bodenheimer, P., Burkert, A., Klein, R. I., \& Boss, A.~P. 2000, in
   Protostars and Planets IV, eds.\ V. Mannings, A.~P. Boss, \& S.~S
   Russell (University of Arizona Press, Tucson), p.~675

Boffin, H. M. J., Watkins, S. J., Bhattal, A.~S., Francis, N., \&
   Whitworth, A.~P. 1998, MNRAS, 300, 1189

Bonnell, I. A. 2000, in Star Formation from the Small to the Large
   Scale, 33rd ESLAB Symposium, eds.\ F. Favata, A.~A. Kaas, \& A.
   Wilson (ESA, Noordwijk; ESA SP-44), in press

Bonnell, I. A. 2001, in The Formation of Binary Stars, IAU Symp.\ 200,
   eds.\ R.~D. Mathieu \& H. Zinnecker (ASP, San Francisco), in press
   (this volume) 

Bonnell, I., \& Bastien, P. 1992, ApJ, 401, L31

Bonnell, I. A., Bate, M. R., \& Zinnecker, H. 1998, MNRAS, 298, 93

Boss, A. P. 1984, MNRAS, 209, 543

Boss, A. P. 1988, Comments Astrophys., 12, 169

Boss, A. P. 1993, in The Realm of Interacting Binary Stars, eds.\ J.
   Sahade, G.~E. McCluskey, \& Y. Kondo (Kluwer, Dordrecht), p.~355

Boss, A. P. 2001, in The Formation of Binary Stars, IAU Symp.\ 200,
   eds.\ R.~D. Mathieu \& H. Zinnecker (ASP, San Francisco), in press
   (this volume)

Burkert, A., Bate, M. R., \& Bodenheimer, P. 1997, MNRAS, 289, 497

Duquennoy, A., \& Mayor, M. 1991, A\&A, 248, 485

Fischer, D. A., \& Marcy, G. W. 1992, ApJ, 396, 178

Goodman, A. A., Benson, P. J., Fuller, G. A., \& Myers, P.~C. 1993, ApJ,
   406, 528

Griffin, R. F. 1992, in Complementary Approaches to Double and Multiple
   Star Research, IAU Colloq.\ 135, eds.\ H.~A. McAlister \& W.~I.
   Hartkopf, ASP Conference Series Vol.\ 32, p.~98

Haraguchi, K., Boffin, H. M. J., \& Matsuda, T. 1999, in Star Formation
   1999, ed.\ T. Nakamoto (Nobeyama Radio Observatory, Nobeyama), p.~241

Heacox, W. D. 1998, AJ, 115, 325

Heacox, W. D. 2000, in Birth and Evolution of Binary Stars, poster
   proceedings of IAU Symp.\ 200, eds.\ B. Reipurth \& H. Zinnecker
   (AIP, Potsdam), p.~208

Heggie, D. C. 1975, MNRAS, 173, 729

Heintz, W. D. 1969, JRASC, 63, 275

Heintz, W. D. 1978, Double Stars (Reidel, Dordrecht)

Heller, C. H. 1993, ApJ, 408, 337

Heller, C. H. 1995, ApJ, 455, 252

Herbig, G. H., \& Terndrup, D. M. 1986, ApJ, 307, 609

Holmberg, E. 1975, in Galaxies and the Universe, eds.\ A. Sandage, M.
   Sandage, \& J. Kristian (University of Chicago Press, Chicago),
   p.~123

Hubble, E. P. 1930, ApJ, 71, 231

Katz, N. 1991, ApJ, 368, 325

Klein, R. I. 2001, in The Formation of Binary Stars, IAU Symp.\ 200,
   eds.\ R.~D. Mathieu \& H. Zinnecker (ASP, San Francisco), in press
   (this volume)

Kormendy, J. 1990, in Dynamics and Interactions of Galaxies, ed.\ R.
   Wielen (Springer-Verlag, Berlin), p.~499

Larson, R. B. 1972, MNRAS, 156, 437

Larson, R. B. 1978, MNRAS, 184, 69

Larson, R. B. 1984, MNRAS, 206, 197

Larson, R. B. 1989, in The Formation and Evolution of Planetary Systems,
   eds.\ H.~A. Weaver \& L. Danly (Cambridge University Press,
   Cambridge), p.~31

Larson, R. B. 1990a, MNRAS, 243, 588

Larson, R. B. 1990b, in Physical Processes in Fragmentation and Star
   Formation, eds.\ R. Capuzzo-Dolcetta, C. Chiosi, \& A. Di Fazio
   (Kluwer, Dordrecht), p.~ 389 

Larson, R. B. 1990c, PASP, 102, 709

Larson, R. B. 1994, in Mass-Transfer Induced Activity in Galaxies, ed.\
   I. Shlosman (Cambridge University Press, Cambridge), p.~489

Larson, R. B. 1995, MNRAS, 272, 213

Larson, R. B. 1997, in Structure and Evolution of Stellar Systems, eds.\
   T.~A. Agekian, A.~A. M\"ull\"ari, \& V.~V. Orlov (St.\ Petersburg
   University Press, St.\ Petersburg), p.~48

Larson, R. B. 1999, in Star Formation 1999, ed.\ T. Nakamoto (Nobeyama
   Radio Observatory, Nobeyama), p.~336

Latham, D. W. 2001, in The Formation of Binary Stars, IAU Symp.\ 200,
   eds.\ R.~D. Mathieu \& H. Zinnecker (ASP, San Francisco), in press
   (this volume)

Lynden-Bell, D., \& Kalnajs, A. J. 1972, MNRAS, 157, 1

Makita, M., Miyawaki, K., \& Matsuda, T. 2000, MNRAS, in press

Mason, B. D., Gies, D. R., Hartkopf, W. I., Bagnuolo, W.~G., ten
   Brummelaar, T., \& McAlister, H.~A. 1998, AJ, 115, 821

Mathieu, R. D. 1994, ARA\&A, 32, 465

Mathieu, R. D. 2001, in The Formation of Binary Stars, IAU Symp.\ 200,
   eds.\ R.~D. Mathieu \& H. Zinnecker (ASP, San Francisco), in press
   (this volume)

Mathieu, R. D., Ghez, A. M., Jensen, E. L. N., \& Simon, M. 2000, in
   Protostars and Planets IV, eds.\ V. Mannings, A.~P. Boss, \& S.~S
   Russell (University of Arizona Press, Tucson), p.~703

Matsuda, T., Makita, M., Fujiwara, H., Nagae, T., Haraguchi, K.,
   Hayashi, E., \& Boffin, H.~M.~J. 2000, AP\&SS, in press

Mayor, M. 2001, in The Formation of Binary Stars, IAU Symp.\ 200, eds.\
   R.~D. Mathieu \& H. Zinnecker (ASP, San Francisco), in press (this
   volume)

Mayor, M., Duquennoy, A., Halbwachs, J.-L., \& Mermilliod, J.-C. 1992,
   in IAU Colloq.\ 135, Complementary Approaches to Double and Multiple
   Star Research, eds.\ H.~A. McAlister \& W.~I. Hartkopf, ASP
   Conference Series Vol.\ 32, p.~73

Moffat, A. F. J., Drissen, L., \& Shara, M. M. 1994, ApJ, 436, 183

Nelson, A. 2000, in Birth and Evolution of Binary Stars, poster
   proceedings of IAU Symp.\ 200, eds.\ B. Reipurth \& H. Zinnecker
   (AIP, Potsdam), p.~205

Nelson, A. F., Benz, W., \& Ruzmaikina, T. V. 2000, ApJ, 529, 357

Pfalzner, S., Henning, Th., \& Kley, W. 2000, in Birth and Evolution of
   Binary Stars, poster proceedings of IAU Symp.\ 200, eds.\ B. Reipurth
   \& H. Zinnecker (AIP, Potsdam), p.~193

Preibisch, T., Balega, Y., Hofmann, K.-H., Weigelt, G., \& Zinnecker, H.
   1999, NewA, 4, 531

Preibisch, T. 2001, in The Formation of Binary Stars, IAU Symp.\ 200,
   eds.\ R.~D. Mathieu \& H. Zinnecker (ASP, San Francisco), in press
   (this volume)

Reipurth, B. 2001, in The Formation of Binary Stars, IAU Symp.\ 200,
   eds.\ R.~D. Mathieu \& H. Zinnecker (ASP, San Francisco), in press
   (this volume)

R\'o\.zyczka, M., \& Spruit, H. C. 1993, ApJ, 417, 677
 
Savonije, G. J., Papaloizou, J. C. B., \& Lin, D. N. C. 1994, MNRAS,
   268, 13

Sawada, K., Matsuda, T., \& Hachisu, I. 1986, MNRAS, 219, 75

Sawada, K., Matsuda, T., Inoue, M., \& Hachisu, I. 1987, MNRAS, 224, 307

Schombert, J. M. 1987, ApJS, 64, 643

Shu, F. H. 1976, in Structure and Evolution of Close Binary Systems, IAU
   Symp.\ 73, eds.\ P. Eggleton, S. Mitton, \& J. Whelan (Reidel,
   Dordrecht), p.~253

Simon, M. 1997, ApJ, 482, L81

Simon, M., Ghez, A. M., Leinert, Ch., Cassar, L., Chen, W.~P., Howell,
   R.~R., Jameson, R.~F., Matthews, K., Neugebauer, G., \& Richichi, A.
   1995, ApJ, 443, 625

Spruit, H. C. 1987, A\&A, 184, 173

Spruit, H. C., Matsuda, T., Inoue, M., \& Sawada, K. 1987, MNRAS, 229,
   517

Stahler, S. W. 2000, in Star Formation from the Small to the Large
   Scale, 33rd ESLAB Symposium, eds.\ F. Favata, A.~A. Kaas, \& A.
   Wilson (ESA, Noordwijk; ESA SP-44), in press

Stahler, S. W., Palla, F., \& Ho, P. T. P. 2000, in Protostars and
    Planets IV, eds.\ V. Mannings, A.~P. Boss, \& S.~S. Russell
    (University of Arizona Press, Tucson), p.~327

Stepinski, T. F., \& Black, D. C. 2000a, A\&A, 356, 903

Stepinski, T. F., \& Black, D. C. 2000b, in Birth and Evolution of
   Binary Stars, poster proceedings of IAU Symp.\ 200, eds.\ B. Reipurth
   \& H. Zinnecker (AIP, Potsdam), p.~167 

Sterzik, M. F., \& Durisen, R. H. 1998, A\&A, 339, 95

Sterzik, M. F., \& Durisen, R. H. 1999, in Star Formation 1999, ed.\ T.
   Nakamoto (Nobeyama Radio Observatory, Nobeyama), p.~387 

Tokovinin, A. A. 1992, A\&A, 256, 121

van Albada, T. S. 1982, MNRAS, 201, 939

Villumsen, J. V. 1982, MNRAS, 199, 493

White, S. D. M. 1978, MNRAS, 184, 185

Whitworth, A. 2001, in The Formation of Binary Stars, IAU Symp.\ 200,
   eds.\ R.~D. Mathieu \& H. Zinnecker (ASP, San Francisco), in press
   (this volume)

Zinnecker, H., McCaughrean, M. J., \& Wilking, B. A. 1993, in Protostars
   and Planets III, eds.\ E.~H. Levy \& J.~I. Lunine (University of
   Arizona Press, Tucson), p.~429

}
\bye